\documentclass[11pt,a4paper,twocolumn]{article}

\usepackage[margin=1in]{geometry}

\usepackage[utf8]{inputenc}
\usepackage[T1]{fontenc}
\usepackage[english]{babel}
\usepackage[final,activate=true,tracking=true,kerning=true,spacing=true,factor=1100,stretch=10,shrink=10,expansion=false]{microtype}\microtypecontext{spacing=nonfrench}

\usepackage{amsmath,amsfonts,amssymb}

\usepackage{graphicx}
\usepackage{xcolor}
\usepackage{tikz}

\usepackage{booktabs}
\usepackage{multirow}
\usepackage{tabularx}

\usepackage{float}     

\usepackage{caption}

\usepackage{authblk}

\usepackage[colorlinks=true,allcolors=blue,breaklinks=true]{hyperref}
\usepackage{url}

\usepackage{comment}

\usepackage{listings}

\usepackage[round,authoryear]{natbib}
\let\cite\citep  

\definecolor{codegreen}{rgb}{0,0.6,0}
\definecolor{codegray}{rgb}{0.5,0.5,0.5}
\definecolor{codepurple}{rgb}{0.58,0,0.82}
\definecolor{backcolour}{rgb}{0.95,0.95,0.92}

\lstdefinestyle{generic}{
  backgroundcolor=\color{backcolour},
  commentstyle=\color{codegreen},
  keywordstyle=\color{magenta},
  numberstyle=\tiny\color{codegray},
  stringstyle=\color{codepurple},
  basicstyle=\ttfamily\footnotesize,
  breakatwhitespace=false,
  breaklines=true,
  captionpos=b,
  keepspaces=true,
  numbers=left,
  numbersep=6pt,
  showspaces=false,
  showstringspaces=false,
  showtabs=false,
  tabsize=2
}
\title{ProMoTA: a model-driven framework for\\
       end-to-end traceability analysis}

\author[$\ast$]{Sadaf Mustafiz}
\author[$\ast$]{Marko Mijalkovic}
\author[$\dagger$]{Moharram Challenger}

\affil[$\ast$]{Toronto Metropolitan University}
\affil[$\dagger$]{University of Antwerp \& Flanders Make Strategic Research Center}

\date{}

\begin{document}
\urlstyle{rm}

\makeatletter
\twocolumn[%
  \begin{@twocolumnfalse}
  \maketitle
  \begin{abstract}
    \noindent
    In this paper, we propose an approach that integrates end-to-end
    traceability with process modelling. Our process models represent
    MDE workflows that span platform-independent modelling,
    platform-specific modelling, and code generation phases. Process
    execution is automated using megamodels and model transformation
    chains. The generation of end-to-end traceability information
    enables global model traceability, from high-level input models to
    generated code, forming the basis for traceability analysis. We
    have built an Eclipse-based framework, ProMoTA, to support our
    approach. ProMoTA extends the Acceleo model transformation
    language, introducing local traceability support. It also includes
    a global traceability map generator and end-to-end traceability
    analysis modules, providing users with a holistic view of the
    entire transformation process. Our framework is demonstrated with
    the use of a Wireless Sensor Network-Based IoT application.
  \end{abstract}
  \bigskip
  \end{@twocolumnfalse}
]
\makeatother

\section{Introduction}
\label{sec:intro}

The landscape of model-driven engineering (MDE) is vast and ever-evolving, with various methodologies and tools emerging to address the multifaceted challenges of the field. Among these, the need for end-to-end traceability has garnered significant attention~\cite{yue2011systematic, leffingwell2002role}. 
The intricate layers of abstraction and subsequent transformations integral to MDE mandate the establishment of a rigorous and comprehensive mechanism to monitor, trace, and validate these transitions. In the absence of such a structured traceability system, the intrinsic complexities associated with multiple transformations can become overwhelming. 
End-to-end traceability is one such technique that aims to help keep complexity under control \cite{garces2017phd}.

Process models provide a detailed road map of the various stages involved in software development or other processes, detailing the tasks, activities, and their dependencies in the workflow \cite{Finkelsteiin1994SPM561316}. These models help in ensuring systematic progression and adherence to the outlined procedures. With a multitude of transformations corresponding to the numerous process steps, maintaining an unambiguous link between original models and their subsequent representations is paramount. This is where traceability steps in. It ensures that every transformation, alteration, or decision taken during the development phase can be traced back to its origin, providing transparency, aiding validation, and ensuring that the final product remains aligned with the initial specifications and intent \cite{aizenbud2006}.

End-to-end traceability stands as a guide in the labyrinth of complex MDE process enactments. It is not merely about connecting the dots; it is about understanding the journey of each dot, each model element, from its inception to its final realization as code. This concept becomes especially pivotal when one considers the multi-layered, often convoluted, transformation processes that models undergo in MDE.

In the early stages of software development, models often represent high-level abstractions, capturing the essence of what a system is intended to achieve \cite{Kent2002}. As the development progresses, these models are transformed, refined, and detailed, eventually leading to executable code \cite{guana2014chaintracker}. Each transformation is a step forward, but it is also a step that can potentially obfuscate the original intent and rationale behind a model element \cite{balajitraceability}. Without a robust traceability mechanism, the lineage of a model element can easily get lost in this transformative journey.

These challenges compound with larger systems. As software systems grow in complexity, the number of model elements and the transformations they undergo also multiply. Suppose you are tracing a single model element through a sequence of several, if not dozens, of transformations. Such a task is daunting, and often not practical to do by hand. Moreover, in the absence of traceability, validating the correctness of transformations, ensuring compliance with requirements, and managing changes become tasks that are often not possible to complete in a meaningful time frame.

End-to-end (E2E) traceability offers a way to navigate such complex scenarios. By providing a comprehensive view of the model elements’ lineage, it elucidates the intricate pathways they traverse. When modelers can see and understand the progression of each model element, they can make more informed decisions during the development process. It also allows stakeholders to ascertain if the transformations align with the original intent, ensuring that the end product meets its desired objectives and specifications \cite{santiago2013}.

In situations where modifications are needed, whether due to changing requirements or detected anomalies, traceability simplifies the identification of impacted elements \cite{needforTR}. This clarity reduces the likelihood of inadvertent errors and the subsequent cost of rectifying them. When teams can pinpoint with precision where changes need to be made, they can do so with efficiency, reducing both time and effort.

Model-to-Model-to-Code (M2M2C) transformation chains represent a progressive transformation chain in modelling that captures the essence of the entire modelling continuum. As the name suggests, this chain embodies a dual-layered transformation - first, a model is transformed into another, more refined model, and subsequently, this refined model is transformed into executable code. While the abbreviation M2M2C suggests a chain of two transformations, the chain can also be longer, as in Model-to-Model-to-Model-to-Code (M2M2M2C) or even more extended chains.
The final transformation in these chains culminates in actual code, which is the tangible, executable manifestation of all preceding models. This code, when executed, brings to life the intentions and objectives first conceived in the initial model, passing through the refinements of the intermediary models \cite{TCFramework}.


Incorporating traceability into these chains ensures that as a model transitions from one phase to another, its integrity, purpose, and objectives are retained \cite{genericTR}. By closely monitoring and documenting these transformations, developers gain a deeper understanding of how each stage contributes to the final code. This transparency offers multiple advantages: it fosters trust in the developmental process, reduces ambiguities, and most crucially, ensures that the final software artefact is a true representation of the initial model's vision \cite{richTrace}.

In the landscape of modern software development, where systems are becoming increasingly complex and the margin for error is shrinking, M2M2C, when combined with E2E traceability, offers a promising modelling framework. It ensures that the journey from conception to execution is smooth, transparent, and in alignment with the system's original goals and objectives.


In this paper, we introduce the \textbf{Pro}cess \textbf{Mo}delling and \textbf{T}raceability \textbf{A}nalysis (ProMoTA) framework for end-to-end traceability analysis of process models. ProMoTA provides support for traceability information generation of local and global traces that form the basis for end-to-end analysis. Tool support has been developed that supports enactment and trace information generation of process models implemented with model-to-model (with ATL) and model-to-code (with Acceleo) transformations. As part of this work, we have extended the Acceleo transformation language with traceability mechanisms. We demonstrate the features of ProMoTA with the use of an Internet of Things (IoT) application. 

The remainder of this paper is structured as follows: Section~\ref{sec:background} provides essential background on model-driven engineering, traceability, process modelling, and megamodelling. Section~\ref{sec:approach} presents an overview of the ProMoTA approach and its architecture. Section~\ref{sec:generation} delves into the generation of end-to-end traceability information, covering both M2M and M2C transformations. Section~\ref{sec:application} describes an example application and its traceability scenarios. Section~\ref{sec:analysis} demonstrates the traceability analysis capabilities of ProMoTA using the example application. Section~\ref{sec:discussion} discusses the limitations of our approach. Section~\ref{sec:relatedwork} reviews related work and finally, Section~\ref{sec:conclusion} concludes the paper and outlines future research directions.
\section{Background}
\label{sec:background}

In this section, we provide essential background on traceability and process modelling. 

\subsection{Traceability in MDE}

Traceability is defined as \emph{the degree to which a relationship can be established between two or more products of the development process, especially products having a predecessor-successor or master-subordinate relationship to one another}~\cite{ieeevoc2010}. 
Traceability is achieved through the maintenance of relationships between model elements and their associated documentation. It is used not just for source code, but also models, documents, data, test results, and more.

The main principle of traceability is about establishing comprehensive connections, named trace links, between correlated elements in various stages of the software development lifecycle. Trace links can be created manually or through the use of automated tools. They serve as the binding threads between related model elements and their associated artefacts.

Manually creating trace links requires engineers to consciously connect related artefacts and express these connections. However, this is generally time consuming and error-prone, and does not scale with larger projects that have many artefacts. Automated tools can streamline the process of creating trace links significantly by automatically detecting related artefacts and creating trace links between them, based on various criteria \cite{uniti2007, maple-sosym2020}.

Trace models are a structured representation of a collection of trace links. As they are models, they can be further consumed by MDE tools and analyzed as part of the MDE software development process \cite{guana2014chaintracker}. Aside from trace links, trace models can also capture additional metadata about the nature and semantics of the relationships they represent. They can include details about the types of artefacts being linked, the original source of the artefacts (such as whether they are outputs of a transformation), and information about how the trace links were generated.

Trace models can be characterized by two key aspects: endogenous and exogenous traceability \cite{andova2012reusable}. Endogenous traceability refers to the tracing of elements within the same model or between models that reside at the same level of abstraction. It primarily concerns the evolution, modification, and refinement of model elements within a given context. As an example, in a class-based design model for a software system, the evolution of a class can be traced across multiple iterations using endogenous traceability. It helps us to understand ``what has been evolved or transformed into what" within a single level of abstraction.

Exogenous traceability relates to tracing model elements across different models that exist at differing levels of abstraction. This is crucial in MDE, where high-level models are transformed via model transformations into low-level models such as analysis artefacts or source code. Exogenous traceability is used to help us understand the derivation paths across differing abstraction layers. As an example, tracing a high-level system requirement to its detailed specification or generated implementation source code, is done using exogenous traceability.


Furthermore, traceability can be classified as pure or tag-based~\cite{vanhooff2007traceability}. Pure traceability relies solely on the existence of a link between the traced entities without any additional information, simply capturing that a connection exists. Meanwhile, tag-based traceability adds more context to the relationship by tagging the trace links with additional information. This can help to specify the nature of the relationship, such as the type of transformation that led to the creation of one artefact from another.

Model traceability can also be differentiated based on the location of the traceability information: extra-model traceability and intra-model traceability~\cite{vanhooff2007traceability}. Extra-model traceability is when trace links are stored outside of the models they trace, usually in a separate trace model or a database. This method allows the creation of complex trace networks that can involve several different models or kinds of artefacts. In contrast, intra-model traceability contains the traceability information within the same model, either as an integral part of the model or as a separate aspect within the same model.

In the context of model transformations, the trace links are between the elements of the associated source and target models. The traceability information generated for each transformation is referred to as \emph{local traceability} or \emph{traceability in the small}. \emph{Global traceability} or {traceability in the large} is captured by connecting the links between the different trace models in the model transformation chain, thus enabling end-to-end navigation throughout a chain of intermediately created trace models~\cite{martes}.

There are many benefits to maintaining traceability in a model-driven development process. Traceability is essential for requirements management. It can help to ensure that all requirements are addressed in a model or section of code, that it is consistent with its design, and that the model can be verified and validated against its operational requirements. Traceability can also help to identify potential errors and inconsistencies in the model \cite{needforTR}. It also contributes to improving the quality of the model by providing a way to track changes to the model over time, ensuring that the model evolves in a controlled and consistent manner. Maintaining traceability can be a challenge, particularly in large and complex projects. However, the benefits of traceability make it an essential part of any model-driven development process \cite{balajitraceability}.

\emph{In this paper, we employ the extra-model and tag-based traceability techniques.}

\subsection{Process Modelling}

\begin{figure*}[tbh!]
    \centering
    \includegraphics[width=\linewidth]{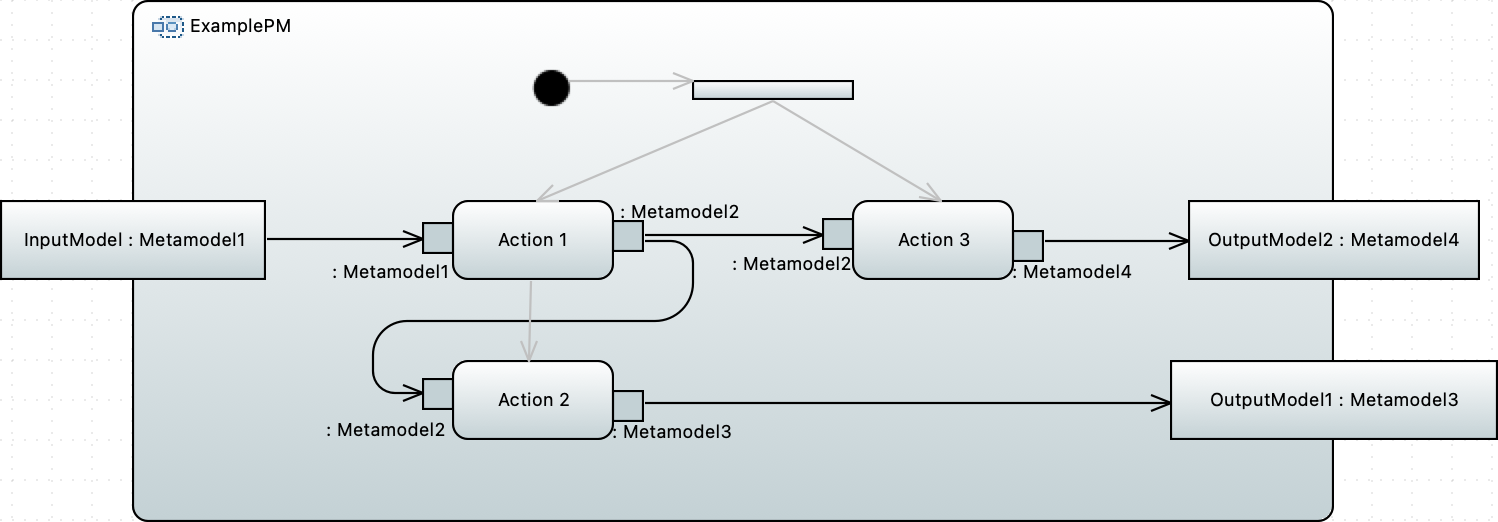}
    \caption{Sample process model.}
    \label{fig:2-process-model-example}
\end{figure*}

Process models are abstract representations of a process or workflow \cite{FeilerSoftwareProcess1992}. They are typically used to model business processes and automated software processes, using languages such as the Business Process Modelling Notation (BPMN)~\cite{bpmn2011}, or UML Activity Diagrams~\cite{umlspec}. They can be used to document and understand processes, analyze processes for bottlenecks or inefficiencies, communicate processes to others, and automate software processes. 

A process model typically includes a number of elements that describe the process. These elements can include the inputs and outputs of the process, the steps involved in the process, the order of the steps, and the conditions or rules that govern the process. The model can also include information about participants or roles associated with each task, and on how the process is monitored or controlled. Process models can be static, meaning they do not change over time, or they can be dynamic, meaning they can be updated as the process changes. Static models are typically used to document and understand fixed processes, while dynamic models can be used when the process is known to change frequently, or there are many variants of the process.

Process models can be used to automate processes, by providing a way to translate the process model into software that can then be executed by a computer, which can either be done with transformations or orchestration software that interprets the process model at enactment time. This can be used to improve processes by identifying bottlenecks or inefficiencies, and by providing a way to communicate process improvements to those who are responsible for the process. Process models can also be used to automate processes that are difficult to automate using traditional methods, such as manual processes or paper-based processes.

Process models can be created via UML 2.0 Activity Diagrams \cite{uml2}. These diagrams have become a preferred choice in modelling business and software procedures due to their versatility and robustness. They offer the capability to concurrently model multiple processes while also efficiently managing their synchronization via fork and join nodes. Referring to Figure \ref{fig:2-process-model-example}, we can see these represented by wide, thin rectangles. Process models generally feature both control and object flow, describing a visual control flow graph that displays how a process can execute, alongside object flow edges, which dictate how data flows through an executing process. An example of object flow can be seen in Figure \ref{fig:2-process-model-example}, with the use of solid black lines. Similarly, control flow is represented by light grey lines. An action node in the model could either signify a basic action, which is a singular step within an activity, or it could depict an entire activity that includes various actions and other activities. Activities in the model can be reused, meaning they can be included in other activity diagrams to invoke certain behaviours. They are denoted with the use of labelled rounded rectangles. To further increase the clarity of the model, the input and output models associated with each activity are explicitly identified via input and output parameter nodes, which are denoted by rectangles on the activity border. UML 2.0 Activity Diagrams have been assigned semantics in terms of Petri Nets \cite{Vitolins}, enhancing their precise formal semantics, and allowing the activity diagrams to be simulated and analyzed.

\subsection{MAPLE-T}

MAPLE-T is an Eclipse-based framework for process modelling, enactment, and analysis \cite{maplet-intent-sam2020, maplet-demo-models2019}. Its main features involve automated enactment of transformation chains, along with traceability extensions that enable semantic analysis of trace models. MAPLE-T is composed of a workspace discovery engine with automated megamodel construction, and a traceability engine that includes support for augmenting transformations with traceability information and generating trace information during process enactments. 

The megamodel~\cite{bezivin2004} is a repository consisting of references to all resources registered in the framework and their relationships. It enables the enactment engine to access resources and execute transformations. The megamodel is aware of process models and all their applicable elements created through the interfaces~\cite{Hebig}, including the transformations and all associated input/output models and metamodels. 
Traceability information generated as a byproduct of the model transformations during process enactment are also retained. During enactment, the megamodel is dynamically updated with all generated artefacts and trace models to include local trace models corresponding to individual transformations in the process model. The megamodel is used to determine which process models can be enacted and how the contained transformations are connected.

MAPLE-T is built on top of the Eclipse Modelling Framework (EMF). 
It enables users to edit process models, register resources in the megamodel, and enact process models. MAPLE-T uses Eclipse Papyrus’ support for UML activity diagrams \cite{papyrus}, which can be used to graphically create and edit process models. 
\section{End-to-end traceability approach}
\label{sec:approach}

The proposed Process Modelling and Traceability Analysis (ProMoTA) framework represents a significant evolution in the field of traceability for Model-Driven Engineering (MDE). While building upon the foundations laid by approaches like MAPLE-T, ProMoTA takes a crucial step forward by addressing the full spectrum of transformations in MDE, from high-level models to executable code.

The key innovation of ProMoTA lies in its comprehensive approach to end-to-end traceability. It extends beyond model-to-model (M2M) transformations to encompass model-to-code (M2C) transformations, thereby bridging a critical gap in the traceability chain. This extension allows for a continuous traceability path from initial abstract models through various transformation stages, all the way to the final generated code.

By incorporating M2C transformations into its traceability framework, ProMoTA offers a more complete picture of the entire model-driven development process. This holistic view is crucial for modern software engineering practices, where understanding the journey from conceptual models to executable code is increasingly important for software maintenance and evolution.

ProMoTA's approach to M2C traceability is pragmatic and efficient. It focuses on element-level tracing, providing a clear link between model elements and their corresponding code segments. This granular traceability allows developers and stakeholders to navigate complex relationships between models and code with greater ease and precision.

\subsection{Architecture}

\begin{figure*}[tbh!]
    \centering
    \includegraphics[width=\linewidth]{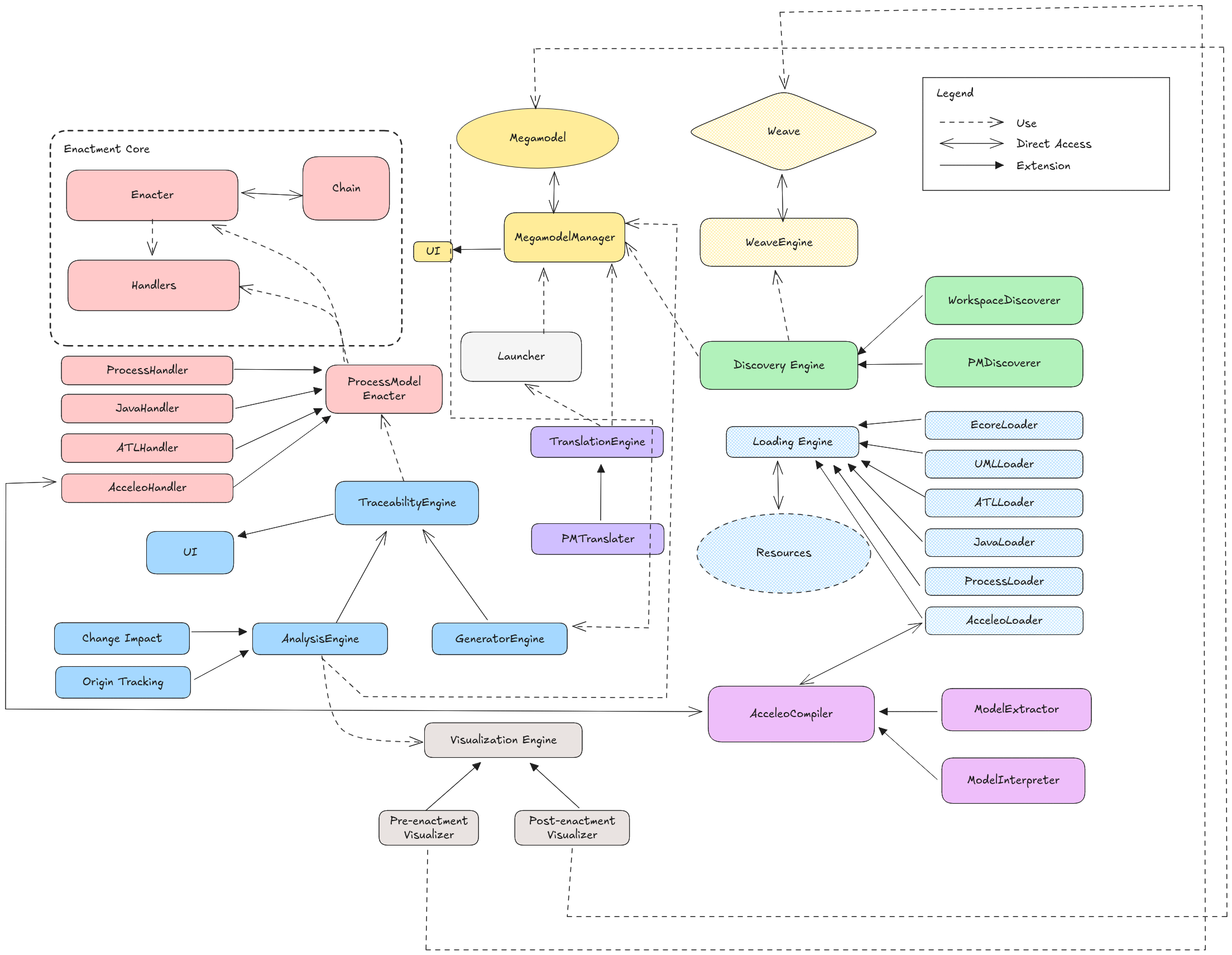}
    \caption{Architecture of the ProMoTA framework [Adapted from \cite{maplet-demo-models2019}.]}
    \label{fig:architecture}
\end{figure*}

ProMoTA is a modular, extensible and EMF-compliant framework built upon the foundations laid by the MAPLE-T approach~\cite{maplet-demo-models2019}. The architecture of ProMoTA, shown in Fig.~\ref{fig:architecture}, is centered on a megamodel - a repository of all pertinent artifacts and their interrelationships within the modelling process.

The megamodel contains references to registered metamodels, transformations, process models, trace models, and other resources fundamental to enactment and traceability analysis. As the process execution progresses, the megamodel is dynamically updated to incorporate traceability metadata resulting from transformation augmentations alongside any newly generated artifacts.

A key component is the automated discovery engine which facilitates the population of the megamodel. Leveraging workspace discovery techniques, it identifies and registers available metamodels and modelling resources. The discovery engine catalogues these resources and their associations within the megamodel, establishing the connections between resources needed to locate and interlink relevant transformations to process models during enactment.

Process execution and workflow coordination capabilities are handled by ProMoTA's enactment engine. Using the megamodel's catalogue of resources, it methodically invokes transformations per process model specifications to automate flows. A major capability is the seamless augmentation of transformations such as ATL (for model-to-model) and Acceleo (for model-to-code) to intrinsically capture fine-grained traceability metadata during execution. This metadata encompasses details including rule names, source and target model elements, and tagged variable references. After execution, trace models are constructed from this metadata and stored within the megamodel.

ProMoTA also features an analysis component that complements the enactment functionality with specialized modules designed to enable common traceability analyses. The analysis modules, such as the Change Impact Analyzer and Origin Tracker, provide valuable insights based on the the extensive catalogue of trace data within the enacted megamodel. The Change Impact Analyzer identifies downstream artifacts affected by alterations to models or transformations. Meanwhile, the Origin Tracker reveals the genesis and lineage of particular generated artifacts and code segments.

Additionally, ProMoTA offers visualization capabilities through an Eclipse-based GUI and editor that allow for graphical creation and analysis of process models. The editor links to the megamodel infrastructure enabling users to visualize the modelling landscape, initiate enactment or perform traceability analysis. Together with a model management layer, these features enable a flexible yet robust modelling environment along with the ability to harness trace data.

ProMoTA provides an integrated, modular architecture combining discovery, enactment, augmentation and analytical capabilities to produce, manage and harness end-to-end traceability across expansive modelling landscapes. The consistent thread interweaving these components is the megamodel, delivering a comprehensive chronicle of artifacts and metadata that proves invaluable for in-depth traceability support within complex modelling environments.

\subsection{ProMoTA approach}

\begin{figure*}[tbh!]
    \centering
    \includegraphics[width=1\linewidth]{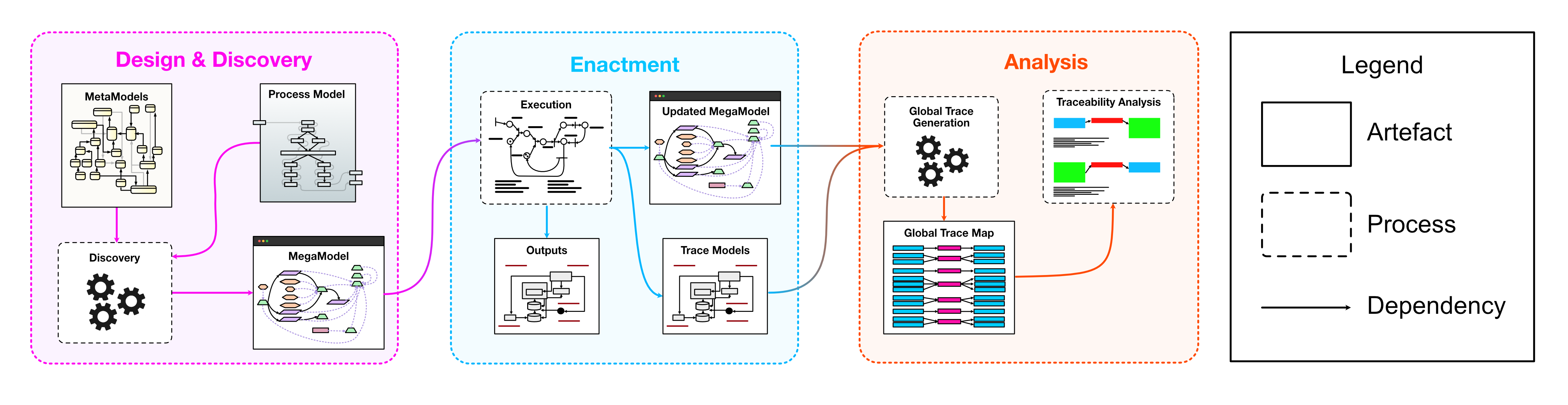}
    \caption{Overview of the ProMoTA approach.}
    \label{fig:promota-approach}
\end{figure*}

ProMoTA is a model-driven engineering framework that provides a comprehensive and flexible solution for the modelling, enactment and traceability analysis of software processes. This approach leverages the benefits of MDE by employing a model-driven approach for managing software development lifecycle processes and associated artifacts. 

A central concept to the ProMoTA approach is the megamodel, a repository of all the artifacts relevant to the enactment and traceability analysis of process models. The megamodel serves as a single source of truth for all models and other associated artifacts, including traceability information collected during process model enactment. ProMoTA uses a megamodel to house registered metamodels, process models, transformations, and trace models. All of these elements are connected through the megamodel, and the different elements can be used in combination to enact and analyze process models.  

Figure \ref{fig:promota-approach} illustrates the ProMoTA approach that covers three main phases: design and discovery, enactment, and analysis. 
As a pre-requisite, PRoMoTA requires the process model to be created using UML Activity Diagrams in Eclipse Papyrus. The models and associated metamodels as well as all transformations should be available in the project.

\textbf{Design and Discovery:} As a prerequisite to this approach, the process model needs to be designed as a UML Activity Diagram in Eclipse Papyrus (see sample in Fig.~\ref{fig:2-process-model-example}). All metamodels as well as the source model instances associated with the process model should be available.

The ProMoTA approach begins with the discovery of all models in the project. Metamodels are first discovered, and then registered in the megamodel. With metamodels registered in the megamodel, process models can also be discovered, with associated transformations being automatically registered and connected. A weave model is used to bind the megamodel and the process model, which is created when a PM is registered in the megamodel. The resulting megamodel provides a set of all relevant model artifacts for a given process model, and connects these elements accordingly.

\textbf{Enactment:} The next step in the ProMoTA approach is process model enactment. The megamodel is used to enact process models using the transformations registered in the megamodel. During enactment, ProMoTA collects trace information which is used to build trace models. Trace models are the result of data gathered during process model enactment, and are used to store all trace information associated with the underlying model transformations used to implement the activities within the process model. The trace models are then stored in the megamodel, and linked to the respective transformations which generated them. The megamodel provides an up-to-date and comprehensive overview of the modelling process. It is a key component of the ProMoTA approach, enabling effective process model traceability and facilitating the enactment and analysis of complex process models.

\textbf{Analysis:} The ProMoTA approach also provides a suite of analyses that operate on the megamodel. These analyses can be used to gain insight into the behaviour of a workflow and its execution via the traceability information stored in the megamodel. 
Additionally, analysis can be carried out to detect anomalies in the trace information, such as discrepancies between the expected and actual output of a transformation. Traceability analysis is performed by reading the megamodel, and creating a \emph{Global Trace Map}, linking all local traces. This map can also be used to visualize the execution of the process model.

ProMoTA allows for the creation of arbitrarily complex process models, suitable for use in systems with numerous activities and transformations. The different elements of the ProMoTA approach can be reused and adapted to different domains, providing a comprehensive end-to-end solution for process model traceability. The megamodel, which is used to retain all relevant models, metamodels, transformations, conformance links, data flow links, and traceability information, can be reused and adapted to different domains. Some implemented analyses have been chosen to support a subset of queries pertaining to specific applications.
\section{End-to-end traceability information generation and analysis with ProMoTA}
\label{sec:generation}

Model-to-Model transformations are pivotal in software engineering, allowing for the generation of software artifacts from high-level models through automated tools, such as the Eclipse Modelling Framework (EMF), the Atlas Transformation Language (ATL), and the Object Management Group’s Query/View/Transformation (QVT) language. The generation of traceability information from transformation executions can be approached in several ways, such as through the use of instrumentation to modify transformation rules, thus capturing the mapping between source and target model elements, or through static analysis of the transformation, applying the results to output models to infer a mapping between source and target model elements. ProMoTA specifically employs the approach of instrumenting existing M2M and M2C transformations to generate traceability models using a higher-order transformation (HOT), providing fine-grained traceability information that allows navigation between input and output models at a granular level.



\subsection{ATL traceability information generation}
\label{subsec:atl_trace_gen}



In the ProMoTA approach, M2M transformations are supported by ATL\cite{atl}, a popular EMF-compliant model transformation language aligned with OMG's Model-Driven Architecture standards. The ProMoTA augmentation transformation is implemented as an ATL transformation, which takes the original transformation as input and outputs a new transformation that includes traceability metadata. 
The metadata encompasses information such as the name of the rule that generated the output model element. This new transformation can be executed similarly to the original transformation, and the generated traceability models can be used to navigate between the input and output models.  

There are several approaches that can be used to generate traceability information from transformation executions. One such approach is the use of instrumentation to modify transformation rules such that they capture the mapping between source and target model elements. Another approach is to statically analyze the transformation and apply the results of this analysis to output models, in order to infer a mapping between source and target model elements. Both of these approaches have their own advantages and disadvantages, and as such the choice of approach depends on the requirements specific to the relevant software development processes.

ProMoTA takes the approach of instrumenting existing M2M transformations in order to generate traceability models, with the use of a HOT. The advantage of this method is that it provides fine-grained traceability information, which can be used to navigate between input and output models at a granular level. This approach is generally applicable to any kind of transformation, and as such is adaptable to various different software engineering processes. The augmentation approach involves modifying transformation rules to capture the mapping between input and output model elements. This is achieved by modifying existing rules to include metadata pertaining to the relevant section of the original transformation that was responsible for generating an output model element.


The generated trace information is stored alongside the output model in a separate trace model. This trace model captures the mapping between the input and output model elements, and can be used to navigate between the two models. The trace model can be queried to obtain information such as the input elements that were used to generate a particular output element, or the output elements that were generated from a particular input element.

\subsection{Acceleo traceability information generation}

Acceleo transformations are a type of model-to-text (M2T) transformation, adhering to the MOFM2T specification~\cite{mofm2t2008omg} and compliant with EMF. They take an input model and produce a specific piece of text as their output. The text output is traced using a text span, which provides information about the location of the text in the output model file. This allows Acceleo transformations to operate on specific elements in the input model and to produce corresponding text in the output model file. Moreover, Acceleo transformations generate trace models, providing a comprehensive representation of the trace information.

ProMoTA supports the use of Acceleo transformations for M2T transformations. In Section \ref{subsec:atl_trace_gen}, we discussed the generation of local traceabilty models from M2M transformations. In this section, we explore how to generate local traceability models from M2T transformations written in the Acceleo language.

\begin{figure*}[tbh!]
    \centering
    \includegraphics[width=\linewidth]{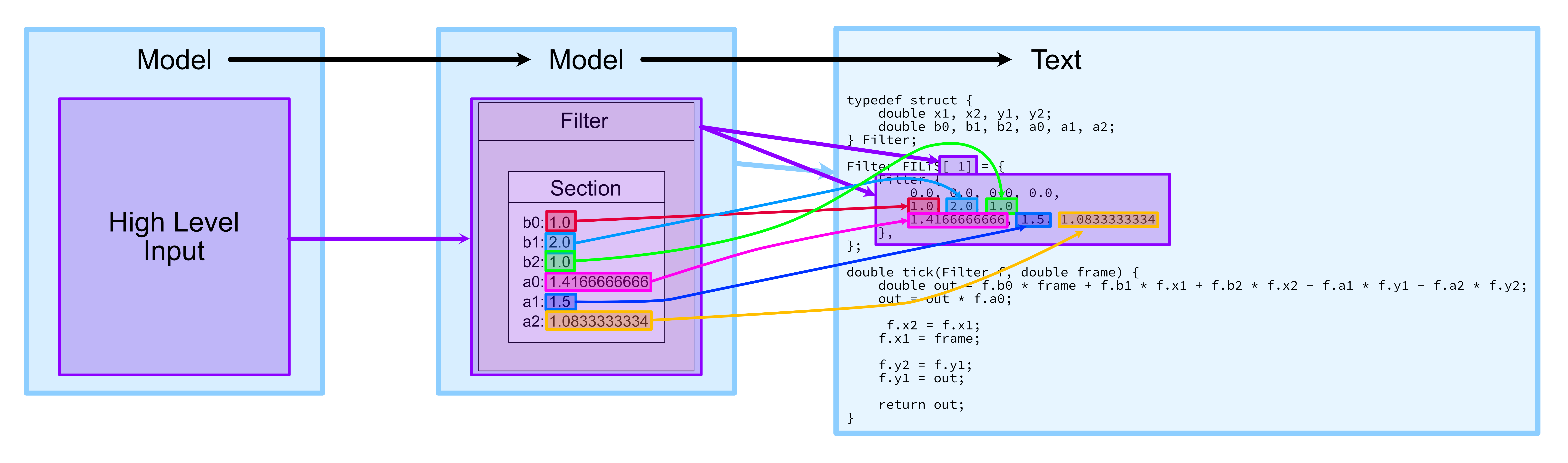}
    \caption{Sample traceability diagram.}
    \label{fig:5-traceability-diagram}
\end{figure*}

Figure \ref{fig:5-traceability-diagram} illustrates the relationship between model elements, such as classes, attributes, and operations, and their corresponding code artefacts in the generated output. In this example, we have an example high-level input model which is transformed into a secondary intermediate model. The specifics of the transformation are not important here, so instead we capture a single element in the input model, represented with a purple rectangle. By following the purple links, we can trace to see how this element ends up in the output text. Additionally, in the intermediate model, we can see new attributes that are established, and how they end up in the output text.

The Acceleo implementation does not support automatic generation of traceability models from transformation executions \cite{balajitraceability}. To address this limitation, we propose an approach to augment Acceleo transformations based on HOTs, similar to how M2M transformations are augmented. In our case, the input transformation is an Acceleo transformation, and the output transformation is a new Acceleo transformation, augmented with traceability information related to the source model.
The transformations are augmented by adding special markers, termed identifier paths, into the output text. They are especially crafted in order to preserve the original intent of the Acceleo code, as well as provide the necessary information for recreating trace models from the output. These paths are used to identify the source model elements that were used in the transformation process, and to construct the trace models from the output text. The original transformation output is also preserved, and can be easily extracted from the augmented transformation output as well.

Our augmentation process first identifies syntactic constructs in the original Acceleo transformation, such as loops and conditionals, and parses the transformation syntax. With this transformation structure identified, we then insert the identifier paths into appropriate locations. These paths contain the necessary information to link the output text to the source model, and to reconstruct the trace models in a post-processing step. Identifier paths are constructed using a series of elements that identify the location of a model element in the source model. They are composed of a series of segments, separated by a delimiter character. Each segment represents a structural element in the source model, such as a class or element. For example, an identifier path for an element might look like the following:

\begin{verbatim}
    filters[0].sections[1]
\end{verbatim}

Here, the segments represent the full path of an element located in a trace link between a source model and the target code artifact. The first segment ``filters[0]" refers to the element at index 0 of a collection called ``filters" in the source model element. The second segment ``sections[1]" refers to the element at index 1 of a collection called ``sections" inside the ``filters[0]" element.

Constructing identifier paths is automated by augmenting the Acceleo transformation template. The template is modified to include additional code that generates identifier paths for each transformed model element. These generated paths are emitted into the output text directly, and retrieved later on a second pass.
With the use of an augmentation transformation, the Acceleo template is modified to output identifier paths directly into the generated code. This involves modifying the template to add code that constructs the identifier path for each transformed model element, ultimately inserting it into the generated code. These generated identifier paths are surrounded by double curly braces \texttt{\{\{like so\}\}}, in order to differentiate them from the rest of the generated source code.

\subsubsection{Augmenting loops}

Loops are a common construct in Acceleo transformations. They are used to iterate over a collection of elements and generate output for each element. Each loop iteration generates a traceable element in the target model. To generate traceability information for these elements, we augment the loop construct with a traceability annotation, which takes the form of an identifier path. This identifier path represents the location of the current element in the source model, and it is constructed using the same method as described above for individual model elements.

To augment a loop with traceability information, we utilize Acceleo's loop index variable \texttt{[i/]} to generate an appropriate index in the identifier path. For example, consider the following Acceleo loop construct:

\begin{lstlisting}[language=java]
[for (e : sourceModel.elements)]
public String [e.name /];
[/for]
\end{lstlisting}

This loop iterates over the elements in the source model and generates a target model element for each one. To generate traceability information for these elements, we can augment the loop construct with a traceability annotation, as follows:

\begin{lstlisting}[language=java]
[for (e : sourceModel.elements)]
    public String {{sourceModel.elements\[[i/]\]}}[e.name /]{{/}};
[/for]
\end{lstlisting}

This annotation indicates that the source model elements are being traced to the target model elements. This information can then be used to trace the source model elements to the target model elements.

\subsubsection{Augmenting conditionals}

Conditionals are another common construct in Acceleo transformations. To generate traceability information for the output generated by conditionals, we can augment the conditional construct with a traceability annotation, which takes the form of an identifier path. This identifier path represents the location of the current element in the source model and is constructed using the same method as described above for individual model elements.

To augment a conditional with traceability information, we utilize Acceleo's built-in boolean variables, such as \texttt{[else/]} and \texttt{[endif/]}, to determine the appropriate location for the identifier path. For example, consider the following Acceleo conditional construct:

\begin{lstlisting}
[if (sourceModel.targetPlatform->size() > 0)]
public String [sourceModel.targetPlatform.name /];
[else]
public boolean improper = TRUE;
[/if]
\end{lstlisting}

This conditional generates output based on whether the source model contains any elements or not. To generate traceability information for the output, we can augment the conditional with a traceability annotation, as follows:

\begin{lstlisting}
[if (sourceModel.targetPlatform->size() > 0)]
public String {{sourceModel.targetPlatform.bane}}
[sourceModel.targetPlatform.name /]{{/}};
[else]
public boolean improper = TRUE;
[/if]
\end{lstlisting}

This annotation indicates that the source model or its elements are being traced to the target model elements generated by the conditional. This information can then be used to trace the source model elements to the target model elements.

By augmenting Acceleo transformation templates with traceability annotations for loops and conditionals, we can automatically generate traceability information for all model elements that are transformed into code artefacts. This information can be used to construct a local traceability map or a global trace map that links all model elements to their corresponding code artefacts.

\subsubsection{Extracting Acceleo trace models}

\begin{figure}[tbh!]
    \centering
    \includegraphics[width=0.35\linewidth]{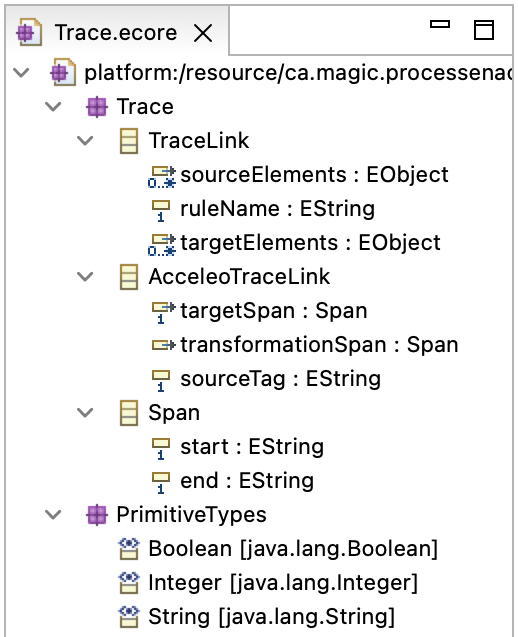}
    \caption{ProMoTA local trace metamodel.}
    \label{fig:ltrace-metamodel}
\end{figure}

After generating identifier paths and traceability annotations in the transformed code, we can extract this information into a separate trace model. Extracting this information allows us to construct a more formal and complete traceability map that links all model elements to their corresponding code artefacts. We can then use this trace model to perform various analysis tasks, such as impact analysis or change propagation.

To extract the traceability information, we can modify the Acceleo transformation templates to generate an intermediate trace model during the transformation process. This trace model captures all the identifier paths and traceability annotations generated in the transformed code. 

The trace identifier tree is a data structure that represents the identifier paths generated during the transformation process. It consists of a root node and a series of child nodes that represent each segment of the identifier path.
Each child node represents a structural element in the source model, such as a class or element. The child nodes are linked together to form a tree structure, where each node represents a level in the identifier path.
For example, consider the identifier path we used earlier:
\begin{verbatim}
filters[0].sections[1]
\end{verbatim}

The corresponding trace identifier tree in Fig.~\ref{fig:4-trace-identifier-tree} contains a branch from the root \texttt{filters} node to a terminating \texttt{sections[1]} element. Child nodes in an identifier tree represent the structural elements of the source model that make up the identifier path.
To construct the trace identifier tree, we read the generated output text and extract the identifier paths and their corresponding code artefacts. Each identifier path is split into its individual segments, and a node is created for each segment. The nodes are then linked together to form a tree structure that represents the identifier path.
For example, consider the following identifier paths:

\begin{verbatim}
filters[0].sections[1]
filters[1].sections[0]
\end{verbatim}

\begin{figure}
\begin{center}
\begin{tikzpicture}[level 1/.style={sibling distance=30mm}, level 2/.style={sibling distance=20mm}]
\node {root}
child { node {filters[0]}
    child { node {sections[1]} }
}
child {
    node {filters[1]} child {
        node {sections[0]}
    }
};
\end{tikzpicture}
\end{center}
\caption{Trace identifier tree.}
\label{fig:4-trace-identifier-tree}
\end{figure}
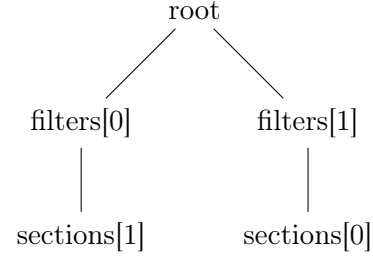

The corresponding trace identifier tree is shown in Fig.~\ref{fig:4-trace-identifier-tree}. In this example, the trace identifier tree has two levels: the first level represents the \texttt{filters} element, and the second level represents the \texttt{sections} element. The leaf nodes of the tree represent the individual filter sections.
Once we have constructed the trace identifier tree, we can use it to construct the global trace map. Our work on the ProMoTA project, including source code, manuals, and tutorials, is available at \url{https://github.com/mde-tmu/promota}.

\subsection{Global trace map generation}

ProMoTA's framework emphasizes the importance of intertwining local trace information to establish a robust traceability infrastructure within process models. The megamodel, serving as the central repository, captures all artifacts and their traceability data throughout the process modelling and enactment lifecycle. The megamodel in ProMoTA is not static but rather evolves, being continuously updated and enriched with new data during the process model enactment. This includes new transformations discovered and enacted, as well as the generation of new trace models that capture the trace information from these transformations. These individual traces are then intricately connected to construct a comprehensive traceability map, offering a full view of the traceability from the initial stages of the process model to its conclusion.



\begin{figure*}[tbh!]
    \centering
    \includegraphics[width=\linewidth]{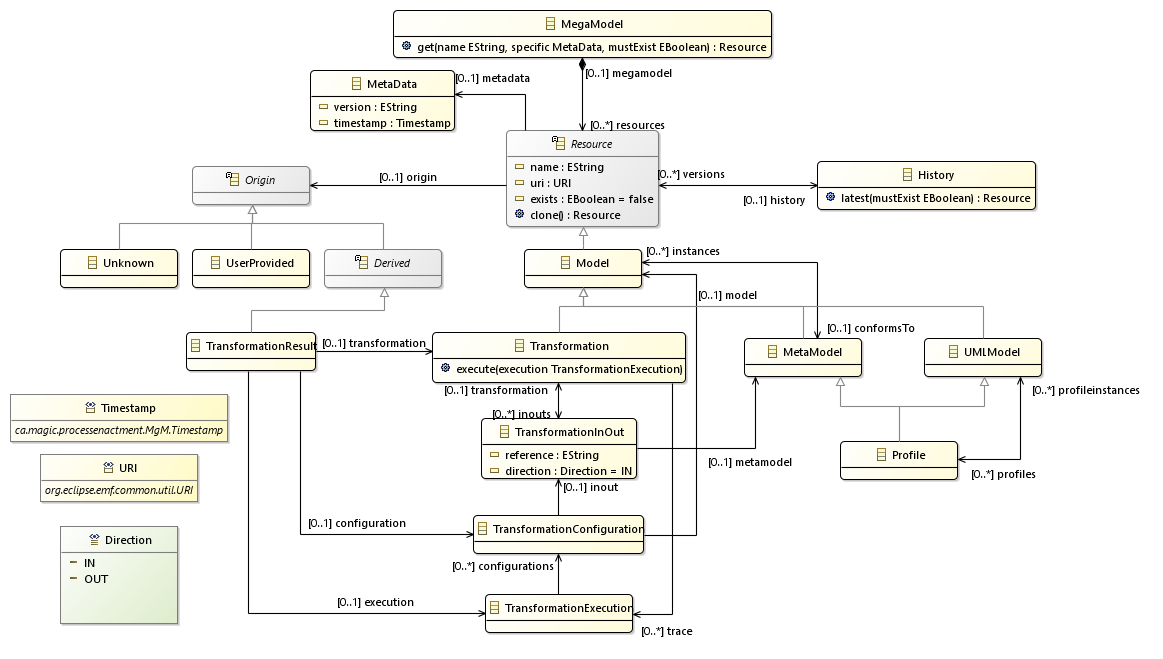}
    \caption{ProMoTA megamodel metamodel.}
    \label{fig:mgm-metamodel}
\end{figure*}

The metamodel of the Megamodel language, adapted from \cite{maplet-intent-sam2020}, is shown in Fig.~\ref{fig:mgm-metamodel}. At the heart of this megamodel is the \texttt{MegaModel} class, which groups together various resources and tracks their changes over time. These resources, denoted by the \texttt{Resource} class, can come from different places: they might be provided by a user, found during a process, or result from transformations. The \texttt{Transformation} class sheds light on how models change and evolve, detailing the configurations and steps involved. The \texttt{Direction} enumeration simply marks whether something is an input or an output in transformations. The megamodel also differentiates between several model types, such as \texttt{UMLModel}, \texttt{MetaModel}, and \texttt{ProcessModel}, each with its own set of features and roles. 

Local trace models generated from both M2M and M2C transformations are stored as distinct resources within the megamodel. Each trace model is assigned a unique identifier and is linked to its corresponding transformation in the megamodel structure. This linkage is implemented using EMF's resource management system, where each trace model is stored as an XMI file and referenced by URI.

The megamodel maintains these trace resources alongside other artifacts such as input models, output models, and transformation definitions. When a transformation is executed, ProMoTA automatically creates a new trace resource, populates it with the generated trace links, and adds it to the megamodel. This approach allows for efficient retrieval and analysis of trace information, as each local trace model can be accessed individually or as part of the larger traceability context.




ProMoTA supports the linking of local trace models to build a comprehensive trace of the process model execution, referred to as the Global Trace Map (GTM). The GTM is a subset of the megamodel providing a comprehensive traceability map that links all model elements to their corresponding code artefacts. It is constructed using the traceability information generated during the transformation process, as well as the trace identifier tree. GTMs provide a holistic view of the system by showing how each model element is linked to its corresponding code artefact and vice versa. 
 
 The linking of local trace models is significant in ProMoTA’s traceability approach, as it allows users to gain a complete understanding of the behaviour of their process models. By linking trace models together, users can observe the relationships between different process models and their component transformations, and can trace the flow of data and transformations across the entire process model. This information can be used to perform various analysis tasks, such as impact analysis or change propagation.

This comprehensive traceability map is a vital component of ProMoTA, enabling users to track the execution of a process model from start to finish. The connection of individual trace models ensures that users can fully comprehend the dynamics of their process models, the interplay between various model components, and the progression of data and transformations throughout the model's lifecycle. Such granular traceability is particularly beneficial in complex domains like Wireless Sensor Networks for Internet of Things applications, where a deep understanding of process models is crucial for pinpointing potential issues and optimizing model performance.


Local trace models generated from both M2M and M2C transformations are stored as separate XMI files in the project workspace. The megamodel maintains references to these trace files through their URIs rather than containing the trace data directly. These are stored as Resources in the megamodel. This URI-based approach keeps the megamodel lightweight while preserving full accessibility to traceability information.

The Global Trace Map is constructed dynamically during process enactment. ProMoTA reads the process model to determine transformation sequences, retrieves trace model URIs from the megamodel, loads the corresponding files, and correlates trace links where outputs of one transformation serve as inputs to another. This ensures the GTM accurately reflects the actual execution path, reading from trace models generated as part of the enactment process.

\subsection{Traceability analysis in ProMoTA}
We discuss here the traceability analysis support provided by ProMoTA.

\subsubsection{Change impact analysis}
Change impact analysis within ProMoTA is an advanced feature that evaluates the effects of modifications to any artifact within the megamodel. By harnessing the traceability information contained within the megamodel, change impact analysis can forecast the ripple effects of changes throughout the entire process model. This analysis is essential for averting the introduction of errors and ensuring that changes do not compromise the system's integrity.

Change impact analysis is initiated by selecting a key element from the input model and employing the comprehensive traceability map to trace all code segments that have a direct or indirect connection to the selected element. The output report generated from this analysis provides an exhaustive list of elements that could be impacted, offering critical insights into the breadth and significance of the proposed changes. This forward-looking analysis is instrumental in upholding the system's quality and dependability as it evolves.

In addition to identifying affected elements, change impact analysis can also help in prioritizing changes by assessing their severity and the effort required to implement them. It can guide decision-making processes by providing a clear picture of the potential trade-offs and risks associated with each change. This level of analysis is particularly useful in large-scale projects where changes can have far-reaching consequences.

\subsubsection{Origin tracking analysis}
Origin tracking analysis is another type of traceability analysis supported by ProMoTA, aimed at uncovering the origins of artifacts and code segments within the megamodel. This feature plays a pivotal role in elucidating the provenance of system components, which is invaluable for tasks such as debugging, validation, and tracking the system's evolution over time.

Origin tracking analysis operates by selecting a specific code section or artifact and meticulously tracing its origins back to the root elements in the model that gave rise to it. This backward tracing process is facilitated by the GTM, which delineates a clear route from the output back to the input. The output report from origin tracking analysis provides a detailed account of the origins of the chosen artifact, shedding light on the developmental journey of each system component from its inception through various transformations.

Origin tracking analysis not only clarifies the genesis of system components but also enhances the understanding of the transformation logic applied throughout the process model. It can reveal patterns in the transformation process and identify opportunities for optimization or refactoring. By maintaining a clear record of the origins of each artifact, ProMoTA ensures that stakeholders can trace back through the development history, simplifying the process of diagnosing issues and understanding the rationale behind the current system state.

\section{WSN-based IoT application: process modelling}
\label{sec:application}

In this section, we present the application domain and use case for demonstrating the features of ProMoTA.  First, the application domain, WSN-based IoT system, is elaborated, and the use case is defined. Next, the process model, languages, activities, and metamodel for the model-driven development of the use case are presented. These artefacts will pave the way for evaluating ProMoTA by analyzing the traceability of the use case in the next section. 

WSN-based IoT systems refer to cooperative and distributed systems that combine Wireless Sensor Networks (WSNs) with the Internet of Things (IoT) technologies. These systems leverage the unique capabilities of both WSNs and IoT to create effective and practical solutions for various applications. They can easily cover a wide area, such as a jungle (to detect a fire) or a farm (to check the need for irrigation). A Wireless Sensor Network (WSN) represents an intricate and decentralized system where the sensor nodes collaborate to form a mesh topology. Within this network, two types of sensor nodes exist: the source and the sink nodes. The source nodes are distributed throughout the medium to gather raw data from their respective environments. This data may encompass a wide array of sensors, such as temperature, humidity, moisture, and air pollution. Each source node is equipped with the capability to communicate with its neighbouring node within a designated range. This neighbour then dutifully transmits the collected data packets to the sink node. Acting as the data output points, the sink nodes aggregate the information gathered by the source nodes and deliver it to higher system layers through a gateway. Usually, the gateway device maintains a physical connection, typically through a serial port, which facilitates the link between a sink node and the system.

The Internet of Things (IoT), on the other hand, is a communication-intensive system that allows embedded devices to have a direct connection to the Internet. By integrating the potential of IoT devices and their frameworks into the distributed topology of WSN nodes, more scalable WSN-based IoT systems emerge, which can have a wide range of applications in nature and industry.
Significantly, the WSN holds a pivotal role within these systems, especially regarding physical area coverage. It serves as the fundamental foundation, mainly when a direct internet connection is unavailable or impractical. In such cases, the IoT nodes come into play and are deployed to specific locations within the environment where an Internet connection can be established. The control of these IoT nodes is managed through Internet-based services, responding to the aggregated data provided by the WSN network. This intelligent synergy between WSN and IoT nodes enables the creation of comprehensive and highly efficient monitoring and data collection systems, which find widespread application in diverse fields and industries.

\subsection{WSN-based IoT system development: process model}

In our previous work \cite{karaduman2021ftg+}, we proposed an FTG+PM (Formalism Transformation Graph and Process Model) for MDE-based development of WSN-based IoT systems using DSML4WSN-IoT. We then adapted our design to provide traceability support for these complex systems, focusing on a portion of the engineering process at the model-to-model transformation level \cite{mijalkovic2022traceability}. 

\begin{figure*}[tbh!]
    \centering
    \includegraphics[width=\linewidth]{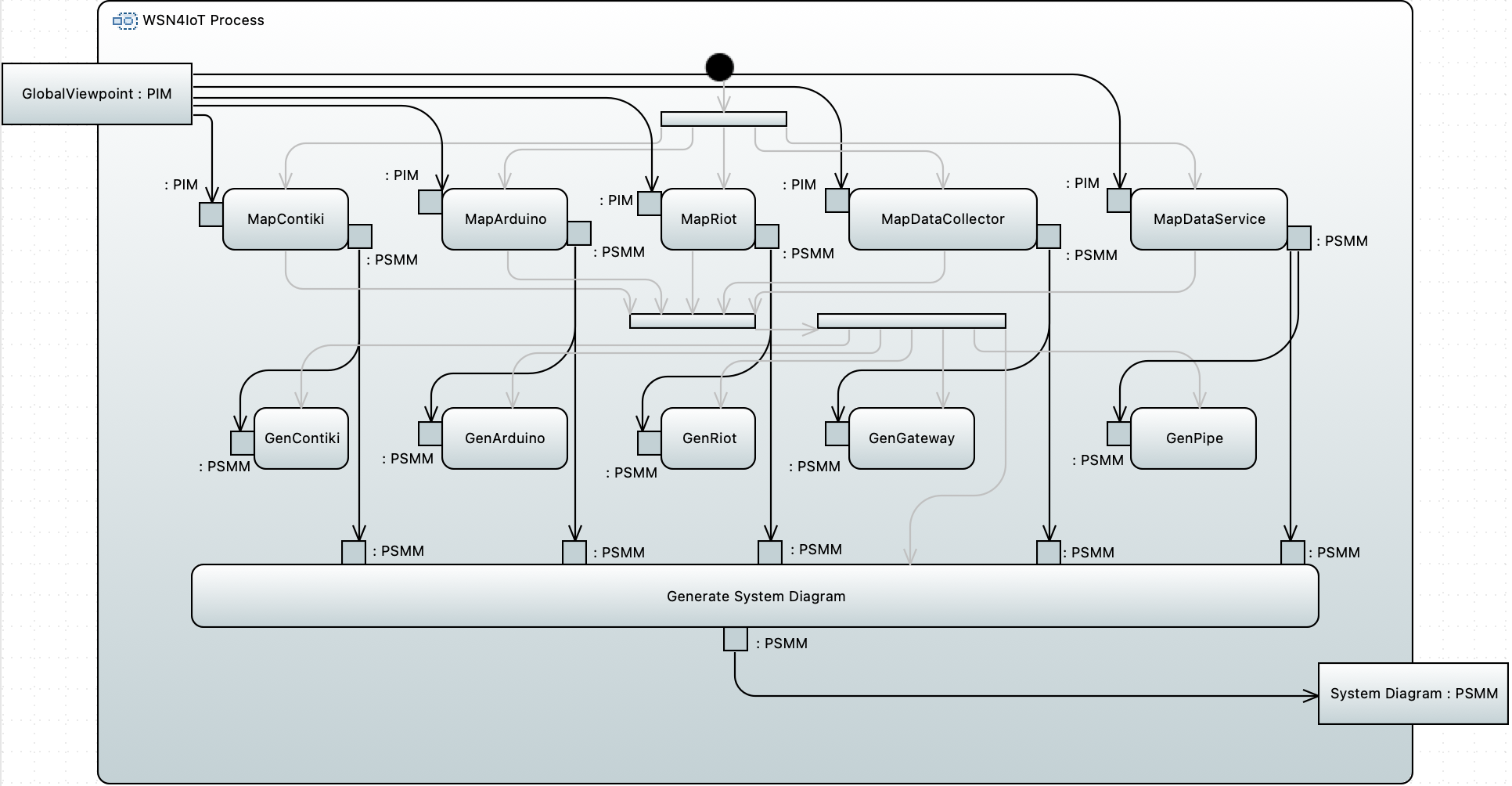}
    \caption{Process model of WSN-IoT application development.}
    \label{fig:process-model}
\end{figure*}

Figure \ref{fig:process-model} presents the updated process model, which builds upon our work in \cite{karaduman2021ftg+}. The process model takes as input a Global Viewpoint, which is a Platform-Independent Model (PIM) conforming to the PIM metamodel. The process can be divided into two main sections: the Model-to-Model (M2M) transformation set and the Model-to-Code (M2C) transformation set.

The M2M portion consists of a series of ATL transformations that transform the Global Viewpoint into various Platform-Specific Models (PSMs), each conforming to the PSM metamodel. 

\begin{itemize}
    \item Map Contiki: Transforms a platform-independent WSN viewpoint model into a platform-specific Contiki Model.
    \item Map Arduino: Transforms an IoT viewpoint model into a platform-specific Arduino Model.
    \item Map RIOT: Transforms an IoT viewpoint model into a platform-specific RIOT Model.
    \item Map Data Collector: Transforms a data collector element into a platform-specific Gateway Model.
    \item Map Data Service: Transforms a data service viewpoint model and the topology viewpoint model into a platform-specific Pipe Model (for a Petri-net model).
\end{itemize}

Following the M2M transformations, the M2C section utilizes a set of Acceleo transformations to transform the generated PSMs into code tailored for each respective platform. 

\begin{itemize}
\item Generate Contiki Code: Generates code for the Contiki platform from the Contiki Model.
\item Generate Arduino Code: Generates code for the Arduino platform from the Arduino Model.
\item Generate RIOT Code: Generates code for the RIOT platform from the RIOT Model.
\item Generate Gateway Code: Generates code for the gateway from the Gateway Model.
\item Generate Node-Red Code: Generates code for the Node-RED platform from the Pipe Model.
\end{itemize}

In addition to the M2C transformations, there is an ATL M2M transformation called \emph{Generate System Diagram} that takes as input all the PSMs generated in the M2M phase and produces a System Diagram. The System Diagram also conforms to the PSM metamodel but provides an overview of the system as a whole. The inclusion of this transformation provides a comprehensive view of the system, facilitating traceability analysis and system understanding. The final outputs of this process include the code generated by the execution of each Acceleo transformation and the System Diagram.

In summary, the updated process model in this paper demonstrates an end-to-end workflow for developing WSN-based IoT systems. It encompasses both M2M and M2C transformations, enabling the generation of platform-specific models and code from platform-independent models. 

\subsection{Languages and activities}

This subsection elaborates on the languages and activities in the process model that will introduce the design views and their transformations to perform an end-to-end engineering flow. Each action in the process corresponds to a manual or automated transformation between different levels of abstraction: PIM, PSM and Code Artifacts.

\textbf{\emph{GlobalViewpoint}}: It is a pivotal representation of the platform-independent model, aligning with the overarching \emph{GlobalViewpoint}. The PIM plays a crucial role in initiating the generation of platform-specific models (PSMs) that cater to specific WSN/IoT platforms. This generation is accomplished through M2M transformations. The process of generating PSMs occurs simultaneously, leading to the creation of various platform-specific models. These models adhere to the \emph{DSML4Contiki}, \emph{DSML4Gateway}, \emph{DSML4RIOT}, \emph{DSML4Arduino}, \emph{DSML4NodeRed}, and \emph{DSML4TinyOS} viewpoints, which represent different perspectives of the PSM language. Each platform-specific model caters to distinct characteristics and requirements of the designated WSN/IoT platforms, enabling the desired implementation.

\textbf{\emph{DSML4Contiki}}: \texttt{Generate Contiki Model} activity takes as input a platform-independent WSN viewpoint model and transforms that model into a platform-specific Contiki Model that conforms to the DSML4Contiki viewpoint. This induced model retains platform-specific elements that belong to ContikiOS \cite{karaduman2018cloud, karaduman2018contikios} such as protothreads, timers process threads, and network.

\textbf{\emph{DSML4Gateway}}: A Gateway model starts with the development of a data collector element at the initial design phase and subsequently associates it with a Gateway Model tailored to the specific platform employed. Within this model, necessary platform-specific component representations of functions, timers, and actions are incorporated. These functions enable the distributed data gathering based on designated time intervals. This also empowers users to trigger specific actions or execute custom code wrapped in functions/methods if required. 

\textbf{\emph{DSML4RIOT}}: In the context of the IoT domain, the DSML4RIOT design process involves transforming an IoT viewpoint model into a RIOT Model \cite{karaduman2020platform}, which is specific to the RIOT operating system. The resulting instance model of a DSML4RIOT comprises platform-specific elements that correspond to various components within the RIOT OS, including timers, sensor/actuator modules, socket libraries, and more. This language formalism enables the interplay and integration of these components, facilitating the development of IoT-focused systems powered by an embedded operating system. In this way, developers can harness the capabilities of RIOT's specialized functionalities to create robust and tailored IoT applications and solutions utilizing DSML4RIOT.

\textbf{\emph{DSML4ESP}}: Creating the Arduino Model involves the conversion of an IoT viewpoint into an Arduino-ESP model \cite{karaduman2021model}. The resulting model is designed specifically for programming an IoT-specific microcontroller using the Arduino framework. It incorporates essential platform-specific elements for the deployment of IoT functionalities, such as Wi-Fi connectivity, various sensors, and actuators. By generating this Arduino model, developers gain a practical tool to effortlessly program and control IoT devices, making the process more straightforward to build IoT applications. This leverages the capabilities of Arduino-based microcontrollers for effective communication, data sensing, and physical actuation tasks.

\textbf{\emph{DSML4Node-Red}}: Forming a NodeRed model involves taking two inputs: the Data Service viewpoint and the Topology viewpoint, which are then transformed into a cohesive NodeRed model. This specialized model incorporates specific elements from Node-RED to facilitate the visual programming aspect of IoT applications. Additionally, the topology model contributes elements required for building mesh topologies of Wireless Sensor Network (WSN) nodes. In this way, developers gain a facilitated design environment to represent the visual programming and representation capabilities for WSN-based IoT systems.

\textbf{\emph{DSML4TinyOS}}: The process of generating the TinyOS model involves converting a WSN viewpoint model into a specialized TinyOS model. This model includes platform-specific elements of the TinyOS operating system, such as tasks, events, and modules, which are essential for crafting a fully functional WSN system.

\textbf{\emph{DSML4SystemDiagram}:} This language provides the System viewpoint in the form of a platform-specific overview diagram, which collectively represents all the platform-specific models. This approach allows the user to observe the platform-specific elements more abstractly and comprehensively, providing a higher-level view of the entire system's structure and components. In this way, the system diagram empowers users to grasp the system's intricacies without getting lost in unnecessary technical details. This birds-eye view of the system enhances clarity and simplifies the understanding of how different platform-specific components interact and contribute to the overall system functionality.

\subsection{Process enactment}

\begin{figure*}[tbh!]
    \centering
    \includegraphics[width=\linewidth]{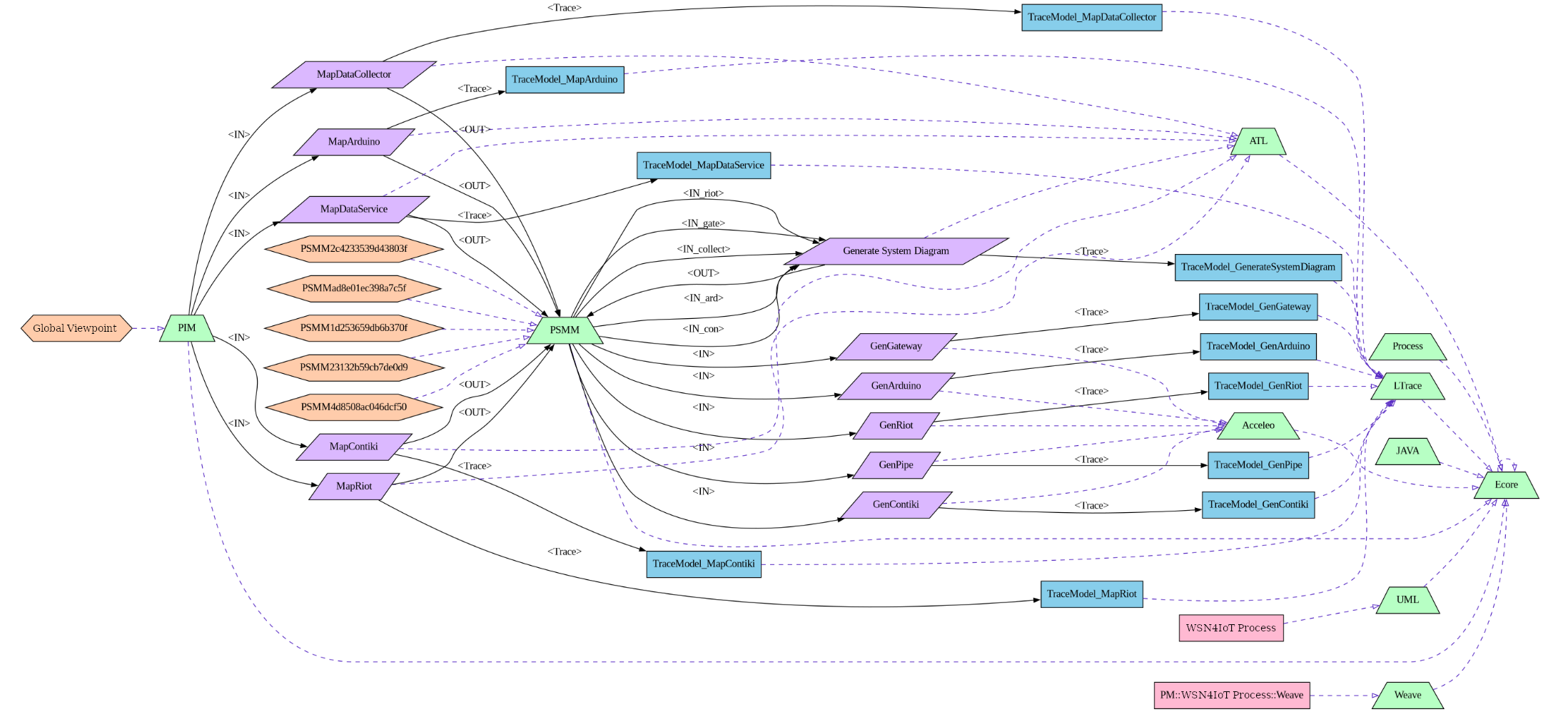}
    \caption{WSN-based IoT application: post-enactment megamodel (purple parallelograms $\rightarrow$ transformations, orange hexagons $\rightarrow$ models; green trapezoids $\rightarrow$ metamodels,  purple dashed arrows $\rightarrow$ conformance links, and solid black arrows $\rightarrow$ transformation data flow).}
    \label{fig:mgm-post}
\end{figure*}

Figure \ref{fig:mgm-post} illustrates the megamodel for the WSN-IoT development process generated after enactment of the process shown in Fig.~\ref{fig:process-model}. This megamodel captures relationships between the artifacts that will be generated during process enactment, including the Global Viewpoint PIM, multiple platform-specific models (Contiki, Arduino, RIOT, and Gateway), alongside generated code for each platform.

The structure shows how the Global Viewpoint serves as input to four M2M transformations (DSML4Riot, DSML4DataCollector, DSML4Contiki, and DSML4Arduino), each producing a platform-specific model tailored to a specific technology, each of which is then served as an input to its respective M2C transformation (Generate\_Riot.java, Generate\_DataCollector.java, Generate\_Contiki.java, and Generate\_ESP.java) to generate executable code.

Generated trace models enable engineers to track how elements in the Global Viewpoint are transformed through the various platform-specific models and ultimately into the generated code, supporting impact analysis and change management across the entire WSN-IoT system.

\section{WSN-based IoT application: traceability analysis}
\label{sec:analysis}

This section discusses the necessity of traceability for WSN-based IoT Systems and two application scenarios. It also provides the global trace map with specific analysis (change impact and origin tracking) for the case study.\\


\noindent \textbf{Scenario 1: IP distribution and modification}\\
In WSN-based IoT systems, the sensor nodes receive a unique IP address according to their gateway. As these systems are distributed and scaleable in a physical environment, their location may require changing, or any specific IP address might need to be assigned after the deployment of the nodes. Therefore, the user/developer has to search and find the specific node's program among the generated code artifacts and find the IP definition scope to replace the target (old) address with the new (candidate) one. Considering the number of nodes, their further scalability, and the software-readaptation after developing the code-base may lead to regression issues, time cost and burden. Particularly, the candidate IP address should be checked against the target IP to ensure it does not cause any IP conflict at design time. Suppose there is no conflict, i.e., the candidate IP is not already assigned. In that case, the corresponding model elements, its code artifact and the scope of the IP assignment inside the code file can be traced and presented to the user without re-exploring the whole model elements and re-synthesizing the whole software. The user then replaces the IP address with the new one and redeploys the code to the node. If there is an IP conflict, the conflicting model elements can be traced automatically and shown to the user to devise an alternative way.\\



\noindent \textbf{Scenario 2: Propagation delay analysis}\\
In another case, the user may design a network topology connecting the node elements to visualize the order of the nodes that will be deployed in the environment. Nevertheless, the topology design is not a straightforward task as the devices are constrained by battery-power, consumption of transmission, and message-sending period as discussed in \citet{karaduman2020analyzing}. This limits the number of neighbour nodes that can be held by a node. Specifically, if a node is responsible for delivering a lot of incoming messages from many nodes, this may create a critical bottleneck for the lifetime of the network, creating a partial fall of a branch. Considering a wide network topology, this may be challenging for the user to pinpoint these nodes and their dependent connections. Alternatively, in order to verify the design, the network topology can be inspected in a formal domain utilizing specific analyses. The bottlenecked node(s) can then be detected and traced to guide the user in devising alternative connections by also presenting the impacted connections both at the model and code artefact levels.

\subsection{Global trace map}

\begin{figure*}[tbh!]
	\centering
	\includegraphics[width=1.0\linewidth]{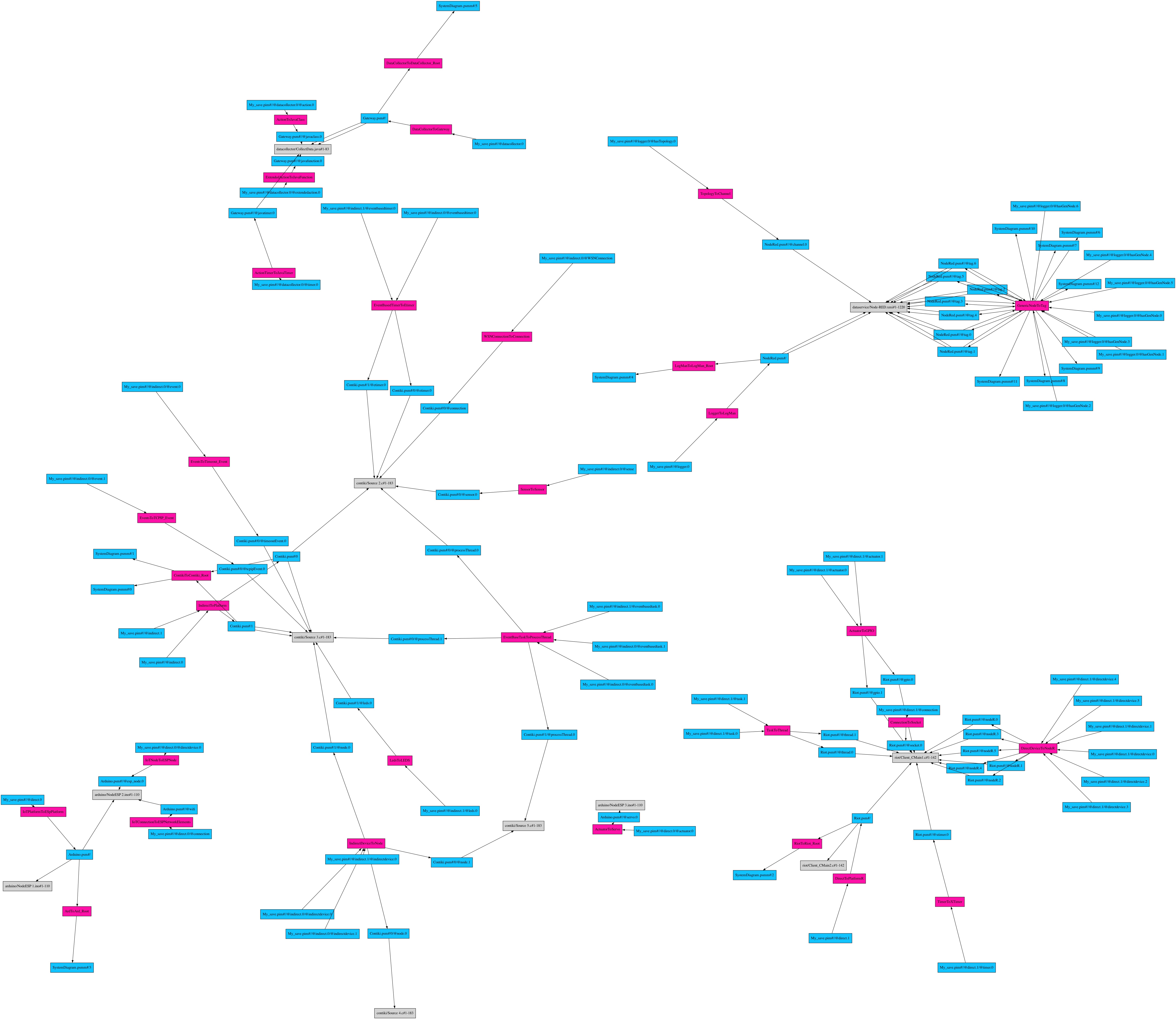}
	\caption{WSN-based IoT application: Global Trace Map (GTM).} 
	\label{fig:6-gtm}
\end{figure*}

The Global Trace Map (GTM) is a subset of the megamodel that provides a consolidated view of the entire design process. As illustrated in Figure \ref{fig:6-gtm}, the GTM acts as a comprehensive visualization tool, offering an exhaustive overview of the WSN-based IoT system development process. It can be thought of as a detailed flowchart or network that not only showcases the different stages in the design sequence but also highlights the critical links between these stages.

The main components of the GTM are the elements and the transformations. In the context of the GTM, elements represent high-level model entities involved in the development process, such as the PIM4WSN, PIM4IoT, and various PSMs (e.g., DSML4Contiki, DSML4Arduino, DSML4RIOT). These elements encapsulate specific data or attributes related to the corresponding models.

Transformations in the GTM encompass both M2M and M2C transformation rules. These rules dictate how a model element evolves, translates into another model element, or translates into specific code during the design process. Rather than representing explicit calculations or individual actions, these transformations capture the overarching rules or guidelines that steer the transition between different models or from model representations to code outputs.

The GTM offers a higher-level view of the WSN-based IoT system development process. Each element in the trace map corresponds to a distinct model component in the design sequence, and the transformations highlight the overarching rules that connect these elements. For instance, a transformation might denote the general rule that guides how a WSN viewpoint model in the PIM is transformed into a platform-specific Contiki model. It would not detail the exact transformation logic but would emphasize the overall transformation rule applied.

This refined perspective ensures that the GTM remains focused on broader design strategies and relationships between main model elements, rather than delving into minute details. Such an approach is beneficial for understanding the overarching design flow, ensuring traceability at the model level, and facilitating discussions and optimizations based on general design structures and rules.

\begin{figure*}[tbh!]
	\centering
	\includegraphics[width=1.0\linewidth]{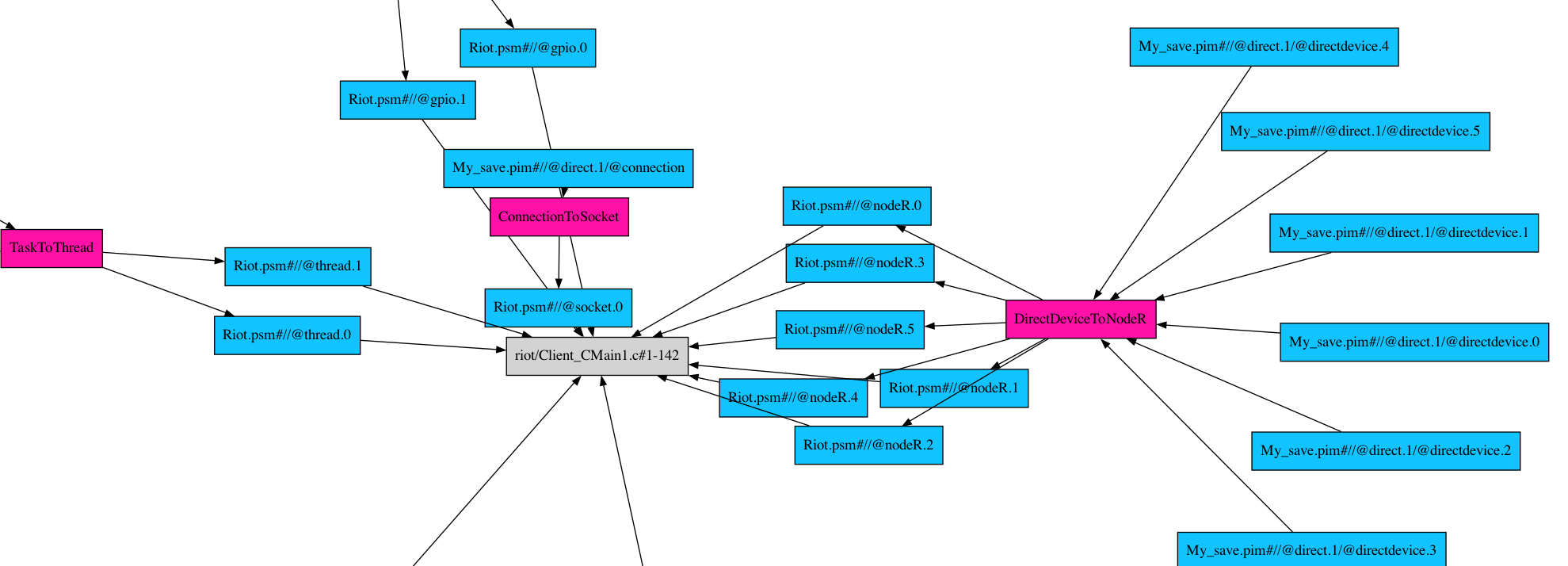}
	\caption{WSN-based IoT application: slice of the Global Trace Map.} 
	\label{fig:6-subsec-gtm}
\end{figure*}

Figure \ref{fig:mgm-post} depicts the megamodel for the WSN-based IoT system application, providing a comprehensive view of all the artifacts involved in the development process. The GTM, as a subset of the megamodel, is specifically displayed in Figure \ref{fig:6-subsec-gtm}. By examining the GTM, stakeholders can gain a clear understanding of the relationships and dependencies between different models and transformations, enabling effective traceability analysis and system comprehension.

\subsection{Change impact analysis}

In WSN-based IoT networks, IP assignment is a critical aspect that determines how devices within the network communicate with each other. In the context of our traceability scenario, the distribution of IP addresses occurs based on the subnet mask defined in the IoT platform model.

Suppose a change is made to the subnet mask value in the IoT platform model. This modification can have significant implications for the IP address assignment and, subsequently, the connectivity of devices within the network. To analyze the change impact, we leverage the ProMoTA framework, which enables us to trace the transformation process and identify the dependencies between the elements.

\begin{figure}[tbh!]
\centering
\includegraphics[width=0.4\linewidth]{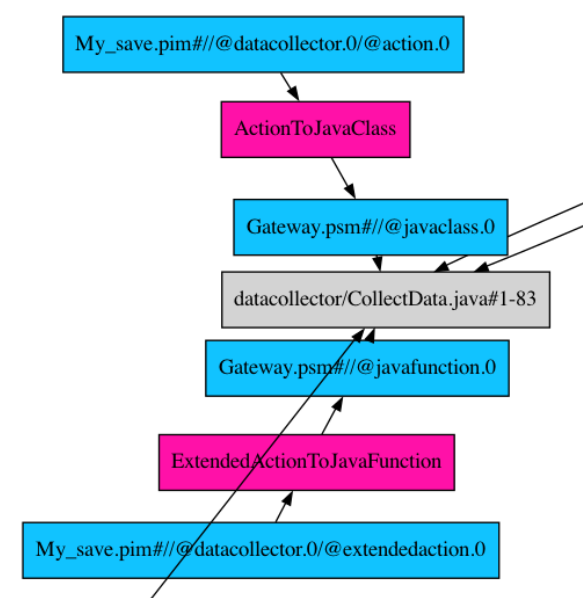}
\caption{Excerpt from global trace map showing gateway-related traceability links.}
\label{fig:gtm-excerpt-gateway}
\end{figure}

Figure \ref{fig:gtm-excerpt-gateway} shows an excerpt from the Global Trace Map that contains relevant traceability information used in change impact analysis. This visualization demonstrates how ProMoTA maintains the relationships between PIM elements (such as Gateway actions), PSM elements (Java classes and functions in the Gateway model), and generated code artefacts. These traceability links form the foundation for analyzing change propagation across the development artefacts.

\begin{figure}[tbh!]
\centering
\begin{lstlisting}[language=C++]
Input.pim:/indirect/#0/indirectdevice/#0
=> ...PSMM:Platform//
==> ...Sink.c:1:181
\end{lstlisting}
\caption{WSN-Based IoT system application: change impact analysis.}
\label{fig:change-impact}
\end{figure}

By traversing the traceability links shown in the Global Trace Map (as exemplified in Figure \ref{fig:gtm-excerpt-gateway}, we can perform precise change impact analysis. For instance, Figure \ref{fig:change-impact} shows a simplified trace path derived from such analysis, revealing that a change to the subnet mask impacts the IP addresses assigned to the IoT nodes in the platform-specific model, as well as the IP address entries in the gateway code.

Understanding the dependencies and impact of changes is crucial in maintaining the consistency and correctness of the network configuration. By having this traceability information available through ProMoTA's Global Trace Map, stakeholders can assess the consequences of modifying network parameters and make informed decisions about the required adjustments.

To address the change impact and ensure the connectivity and stability of the WSN-based IoT network, the IP address assignment process should be carefully reviewed, and all the affected elements should be updated accordingly. Additionally, automated verification tools can be employed to identify inconsistencies and potential errors that might arise due to the change in IP address allocation.

With continuous traceability and analyzability, stakeholders can effectively manage change impact and maintain the integrity of the network configuration, reducing the risk of connectivity issues and enhancing the overall system performance.

\subsection{Origin tracking analysis}

Propagation delay is a critical parameter that affects the performance and reliability of wireless communication in WSN-based IoT networks. It reflects the time it takes for a signal to propagate from the source to the final destination, primarily determined by the distance between the nodes.

In our origin tracking scenario, we explore how changes in the relative distance between nodes affect the propagation delay. By employing the ProMoTA framework, we can trace the origins of these changes, identify the dependencies, and understand how they propagate throughout the system.

\begin{figure}[tbh!]
\centering
\begin{lstlisting}[language=C++]
Sink.c:1:181
=> ...Input.pim:/indirect/#0/indirectdevice/#0
==> ...PSMM:Platform//
\end{lstlisting}
\caption{WSN-Based IoT system application: origin tracking analysis.}
\label{fig:origin-tracking}
\end{figure}

Figure \ref{fig:origin-tracking} illustrates the origin tracking analysis performed on the distances between nodes in the topology viewpoint model. By tracing the dependencies and transformations, we can see how the relative distances between nodes propagate through the system. This information is crucial for calculating the propagation delay accurately, as it directly affects the communication latency and performance of the network.

In this scenario, we can observe that the changes in distances between nodes in the topology viewpoint model impact the propagation delays calculated in the platform-specific models. These distances are used to calculate the relative propagation times in the final code artefacts.

Understanding the origins and dependencies of propagation delays enables stakeholders to evaluate the impact of changes in the network topology accurately. By having traceability information available, they can take necessary actions to ensure optimal communication performance, such as recalculating and updating propagation delays based on the modified distances.

To facilitate this process, automated tools can be utilized to perform calculations and generate updated artefacts based on the traceability and transformation information. By leveraging the ProMoTA framework's capabilities, stakeholders can effectively track the origins of propagation delays and manage the impact of changes on the network performance.

\section{Discussion}
\label{sec:discussion}


\subsection{Limitations}
There are certain constraints and boundaries that define the scope and applicability of our framework. 
ProMoTA's tracing capability is limited to the element level. This means that while it can effectively trace individual elements across transformations, it does not delve into the attribute level of these elements. 

Currently, ProMoTA only provides support for ECore models. 
Regarding transformation language support, ProMoTA is limited in the transformation languages that are natively supported. ProMoTA is designed to support Acceleo for M2C transformations and ATL for M2M transformations. ATL and Acceleo are open-source, actively maintained, and widely adopted with the MDE community. Their integration into the Eclipse ecosystem ensures robust tooling support, and ensures reliability for long-term research and adoption. 


When it comes to process models, ProMoTA exclusively supports Papyrus, relying on its activity diagram support. While Papyrus is a renowned and robust modelling tool, this might pose challenges for projects or researchers who employ other process modelling tools or platforms.

However, ProMoTA is modular and extensible and can be fairly easily extended to support other modelling languages and model transformation languages.

\subsection{Use of large language models}

The ProMoTA framework remains highly relevant in the age of Large Language Models (LLMs) because the task of E2E traceability analysis in model-driven engineering demands formal, structural, and deterministic guarantees that LLMs are not inherently equipped to provide. ProMoTA's fundamental mechanism involves automatically augmenting transformation rules, such as those in ATL (M2M) and Acceleo (M2C), to intrinsically capture fine-grained traceability metadata during execution. LLMs, which operate primarily on linguistic patterns and probabilistic inference, cannot reliably ensure the correctness and completeness required for validating transformations across multi-layered, convoluted MDE processes, where a model element might undergo dozens of transformations and its lineage could easily be lost. The work specifically addresses managing this complexity by storing all artifacts and trace data in a centralized, dynamically updated megamodel and consolidating local traces into a global trace map, enabling vital, non-linguistic analyses like origin tracking and change impact analysis. The need for rigorous methods to manage the increasing complexity of systems, such as IoT and cyber-physical systems, where the margin for error is shrinking reinforced the necessity of ProMoTA's automated, structural approach over general-purposed LLM solutions.  
\section{Related Work}
\label{sec:relatedwork}
 

This section discusses related work on requirements traceability, local and global traceability, and the use megamodelling in traceability analysis. It also covers the state of the art in traceability analysis of IoT and CPS applications.

\subsection{Requirements traceability}
The significance of ensuring robust traceability in system development has been widely recognized in research and industry. Leffingwell and Widrig~\cite{leffingwell2002role} provided foundational work highlighting the crucial role of requirements traceability in system development. Furthermore, Yue et al.~\cite{yue2011systematic} conducted a systematic review emphasizing the need for robust transformation approaches between user requirements and analysis models. The importance of maintaining these links from initial requirements throughout development has been a continuous focus in the field. Surveys show that there is work on traceability but limited work on end-to-end traceability \cite{galvao2007}. 

More recently, specialized applications and tooling have emerged alongside efforts to manage traceability complexity. Moros et al~\cite{moros2013} focused on transforming and tracing reused requirements models to home automation models, specifically addressing requirements evolution. Bouzidi et al~\cite{bouzidi2024} developed a unified traceability method aimed at bridging the gap between business and software engineering. Others have focused on enhancing comprehensive tool support, such as LISSA~\cite{asuncion2007}, an end-to-end industrial software traceability tool. Furthermore, Vogelsang et al.~\cite{vogelsang2025impact} address challenges in traceability automation, such as investigating the impact of requirements smells in prompts used for automated traceability.

Despite these advances, most existing traceability tools offer a narrow lens, focusing either on tracing the lineage within specific model-to-model transformation chains or only providing localized traceability for one particular transformation language \cite{winkler2010}. This piecemeal approach falls short of providing a holistic view of an artefact’s complete evolutionary journey, thereby hindering comprehensive traceability analysis and effective requirements/software management. ProMoTA's core contribution, in contrast, provides comprehensive E2E traceability spanning M2M and M2C transformations across heterogeneous MDE chains, managing all trace data through a persistent global trace map stored within a central megamodel. This specific focus on instrumented M2M2C traceability and structured megamodelling provides the analytical basis necessary for traceability analysis, such as change impact and origin tracking, setting ProoTA apart from systems primarily focused on requirements management or runtime concerns. 

\subsection{Megamodel-Based Traceability Information Generation and Analysis}

Model-to-Model (M2M) traceability is crucial for understanding the transformation process and ensuring the consistency of transformed models. Various researchers have explored this field, introducing distinct methodologies and tools. This section reviews these works, emphasizing their contributions and limitations.

Beyhl et al. \cite{Beyhl2013AMM} introduced a novel framework that maintains traceability links within a hierarchical megamodel. This megamodel amalgamates both high-level and fine-grained artefacts, offering a comprehensive view of the transformation process. However, a notable limitation of their approach is the absence of support for global traceability. Similarly, the MegaMart2 ECSEL project~\cite{megamart} focused on megamodelling at runtime as a scalable, model-based framework designed for continuous development and runtime validation of complex systems. In our case, ProMoTA leverages the megamodel not primarily for runtime concerns, but as a central, persistent repository to explicitly generate, retain, and link local traces. 

\subsection{Local traceability}

Other works have focused on augmenting transformations languages for local tracing. Falleri et al.~\cite{Falleri06c.:towards} presented a framework that augments Kermeta~\cite{kermeta} model transformations with traceability support, but this work requires manual enhancement of the transformations. 
Balaji \cite{balajitraceability} explored adding Acceleo and Xtext support to \emph{Chain Tracker}. 
While facing challenges with Acceleo, they implemented rudimentary support for local traces on M2T transformations, but this implementation only supports Xtext models. 
Oldevik \cite{oldevik2006traceability} proposes a conceptual model for traceability capturing M2T and M2M transformation traces, including source code blocks and positions. 
Garcia et al. \cite{garcia2014testing} also provided a conceptual overview of augmenting Acceleo for local traceability using HandyMoF. 

All of the above work~\cite{Falleri06c.:towards, balajitraceability, oldevik2006traceability, garcia2014testing} only supports local traceability information generation. Model transformation chains with global traceability information are not generated or stored, and in general this work does not examine end-to-end traceability analyses.

Initially, we first tried to use Acceleo's native traceability support to create local M2T trace models. However, this functionality was removed several years ago and no longer exists as part of the Acceleo software distribution. None of the related works cited above create a megamodel with a global trace map, and thus cannot create artefacts suitable for meta-analysis specific to a given modelling domain.

\subsection{Global traceability}
Pilgrim et al. \cite{pilgrim2008constructing} add support to UNiTI for generating trace models during model transformation chain enactment. The support introduced here does not include M2T transformations, and does not specify a method by which trace model generation is automatically added to transformations which are not instrumented with traceability support.

Guana et al. \cite{garces2017phd} proposed Chain Tracker, an environment tailored for generating traceability in model transformation chains, predominantly focused on M2M ATL transformations. While effective, the traces generated by Chain Tracker are confined to the transformation level and do not integrate into a megamodel.
 Additionally, their approach does not support M2C transformations. The tool was not found to be publicly accessible, and seems to be a previous version of the work described in \cite{garces2017phd}. 
Garces \cite{garces2017phd} extended Chain Tracker to support E2E traceability for M2M and M2T (Acceleo) transformations.
However, in this extension, links are stored at the transformation level and are not retained in a globally linked structure like a megamodel, instead joining links dynamically during trace analysis. 
Moreover, unlike their work we use process models created via Papyrus to model a process for enactment and traceability analysis, providing a structured workflow mechanism not explicitly employed in Chain Tracker's analysis approach. 

Hassane et al. \cite{hassane2019maple} introduce MAPLE-T, an extension to MAPLE \cite{mustafiz2018maple} which adds support for traceability information generation, and generation of global trace maps in the context of software process modelling and enactment. These global trace maps enable support for building various new types of traceability analyses based on enactment output.
MAPLE-T serves as a foundational basis for ProMoTA. However, MAPLE-T traceability support is limited to ATL M2M transformations, with no support for Acceleo M2C transformation traceability generation or any M2M2C transformations. MAPLE-T was also tailored for network service management and it's applicability in other domains was not explored.


A recurring theme across these works is the absence of comprehensive traceability that spans both local and global perspectives. Beyhl et al. specifically proposed the use of megamodels to offer a holistic view of the transformation process. These megamodels amalgamate both high-level and fine-grained artefacts, showing the potential for facilitating process enactment. Given this potential, it's surprising that more efforts haven't utilized megamodels for process enactment, highlighting a clear gap. This underscores the need for more comprehensive solutions that can bridge the divide between local and global traceability, ensuring a seamless transformation process.

\subsection{Using LLMs for traceability analysis in MDE}
Recent research has explored leveraging large language models (LLMs) to automate traceability link recovery in model-driven engineering contexts. \cite{Bonner2024} combine vector-based similarity with LLM-assisted reasoning to automatically infer and validate trace links between textual requirements and MBSE artifacts, reducing manual analysis. \cite{Fuchss2025} extract architecture entities directly from software documentation and source code to establish traceability between software architecture descriptions and implementation artifacts. \cite{Ali2024} integrate retrieval-augmented generation (RAG) approach with a graph-based representation of source code, allowing LLMs to establish semantic links between natural language requirements, UML design models, and the corresponding source code. 

Although these approaches generate trace links between specific levels of abstraction (e.g., between requirements and design, or between architecture and code), they typically operate on isolated artifact pairs and lack explicit modelling of transformation chains. Whereas LLM-based methods reply primarily on post-hoc semantic inference, ProMoTA achieves structurally and semantically grounded traceability that is explicitly maintained  throughout the model-driven development process.

While each of the aforementioned works offers valuable insights into traceability, they also highlight the existing gaps in the domain. The challenge remains to develop a solution that not only captures local traceability information but also seamlessly links distant models, providing a holistic view of the entire transformation process.
\section{Conclusion}
\label{sec:conclusion}


This paper presents a model traceability framework for end-to-end traceability for model-based processes. ProMoTA provides support for model-driven traceability information generation and management by addressing the complexities inherent in the M2M and M2C transformations that are part of a process model. It supports the generation of local as well as global traceability information for all models involved in the process, providing a global trace map that is retained in the megamodel. 

We built support for local traceability in model-to-code transformations, using the Acceleo language as a traceability target. While the concept of local Acceleo traceability is not new, the manner in which we implemented and utilized it within ProMoTA was instrumental in achieving the desired end-to-end traceability. Megamodels were used to retain trace information and construct global traceability maps. The global traceability information forms the basis for carrying out traceability analysis.


We have demonstrated our approach with the use of a wireless sensor network-based IoT application to enable origin tracking and change impact analysis for two scenarios. 
As future work, we intend to extend ProMoTA with semantically rich traceability information generation and analysis.



\section*{Acknowledgments}
This work is partly funded by NSERC and Toronto Metropolitan
University. The authors would like to thank Burak Karaduman for his
support in shaping the case study of the WSN-based IoT System.

\bibliographystyle{plainnat}
\bibliography{references}

\section*{About the authors}

\paragraph{Sadaf Mustafiz} is an Associate Professor in the Department
of Computer Science at Toronto Metropolitan University, Canada.
(\href{mailto:sadaf.mustafiz@torontomu.ca}{sadaf.mustafiz@torontomu.ca})

\paragraph{Marko Mijalkovic} is a PhD student in the Department of
Computer Science at Toronto Metropolitan University, Canada.
(\href{mailto:marko.mijalkovic@torontomu.ca}{marko.mijalkovic@torontomu.ca})

\paragraph{Moharram Challenger} is an Associate Professor in the
Department of Computer Science at University of Antwerp, Belgium.
(\href{mailto:moharram.challenger@uantwerpen.be}{moharram.challenger@uantwerpen.be})

\end{document}